# Flexible Micro Thermoelectric Generator based on Electroplated $Bi_{2+x}Te_{3-x}$

Etienne Schwyter, Wulf Glatz, Lukas Durrer, Christofer Hierold
Micro and Nanosystems, Department of Mechanical and Process Engineering,
ETH Zurich, Switzerland

*Abstract* - **We present and discuss the fabrication process and the performance of a flexible micro thermoelectric generator with electroplated $Bi_{2+x}Te_{3-x}$ thermocouples in a SU-8 mold. Demonstrator devices generate $278\mu Wcm^{-2}$ at $\Delta T_{meas}=40K$ across the experimental set up. Based on model calculations, a temperature difference of $\Delta T_G=21.4K$ across the generator is assumed. Due to the flexible design and the chosen generator materials, the performance stays high even for curved contact surfaces. The measurement results correlate well with the model based design optimization predictions.**

## I. INTRODUCTION

Power MEMS are proposed – among other applications – to provide electrical energy for autonomous microsystems and sensors. Micro thermoelectric generators (µTEGs) based on the Seebeck effect convert thermal energy, represented by heat flow ($Q_H^\bullet, Q_C^\bullet$) and driven by a temperature difference across the generator, into an electrical potential difference. For thermoelectric generators the material of choice has to have not just a high Seebeck coefficient but also a good electrical conductivity for a low internal resistance of the generator and a high thermal resistance in comparison to the thermal resistances of the generator's package, which provides the thermal interface to the environment. Bismuth telluride ($Bi_{2+x}Te_{3-x}$) is the state of the art thermoelectric material for power harvesting and cooling applications near room-temperature.

## II. MICRO FABRICATION PROCESS

### A. µTEG Chip Processing

In our previous publication [1] on the fabrication, testing and modelling of fully flexible µTEGs, a first proof of concept was presented. This fabrication concept has been adopted and extended for the use of $Bi_{2+x}Te_{3-x}$ as thermoelectric material, replacing the nickel and copper thermolegs. The fabrication process consists of four deposition, photolithography and etching steps followed by an electrochemical deposition (ECD) of both Bi-rich and Te-rich $Bi_2Te_3$ in 200-500µm thick SU-8 molds, respectively. The chips are mechanically polished and three subsequent deposition, photolithography and etching steps complete the fabrication of the micro thermoelectric generator. For the

$Bi_{2+x}Te_{3-x}$ µTEGs, the material of choice for the electrical interconnects and contacts is gold, since copper would have been oxidized in the electrolyte solution and copper ion diffusion into the $Bi_{2+x}Te_{3-x}$ thermo legs would have deteriorated the generator's performance. The SU-8 mold is left in place and serves as package providing flexibility and stability to the device. The $Bi_{2+x}Te_{3-x}$ µTEG demonstrator device is shown in Fig. 1. Fig. 2 shows the working principle of a generator device.

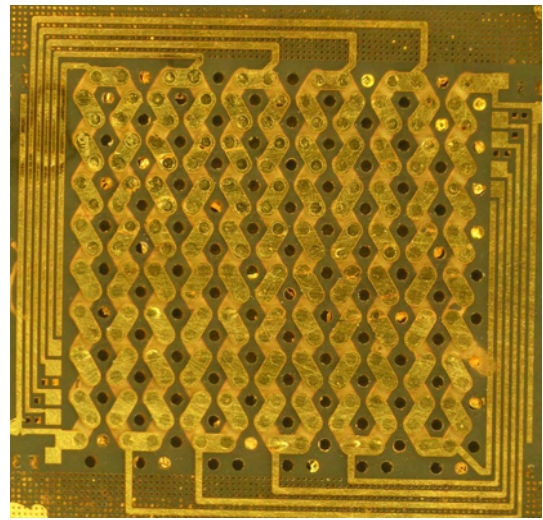

**Fig 1** Micro thermoelectric generator with $Bi_{2+x}Te_{3-x}$ thermolegs and gold interconnects.

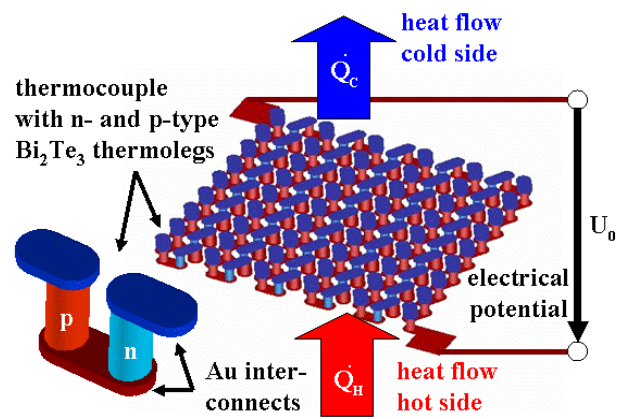

**Fig 2** Working principle of a µTEG connected to a heat source and heat sink.

### B.    Electrochemical Deposition of $Bi_{2+x}Te_{3-x}$

The electrochemical deposition of $Bi_{2+x}Te_{3-x}$ has been studied intensively during the last few years [2,3]. The full control over the stoichiometric ratio between the compounds is the most challenging task for ECD of $Bi_{2+x}Te_{3-x}$. However, electroplating is cheap, fast and efficient, which makes this method interesting for mass production.

The electrolyte solutions are prepared by dissolving 80mmol $O_2Te$ and 20-60mmol $Bi_2O_3$ in 2M $HNO_3$ (65%). Thus a 2M $HNO_3$ solution containing $Bi_3^+$ and $HTeO_2^+$ ions with a measured pH-value of 0.02 is achieved. All depositions are performed with a three electrode configuration, where a Mercury Sulfate Electrode (MSE) is used as reference electrode. A Pt grid is used as counter electrode and the structured SU-8 molds were connected to the working electrode. It is advantageous to apply a pulsed plating method to homogeneously fill molds with high aspect ratios. Primarily the ion concentration at the reaction surface does not deplete and secondly the undesired reaction products have time to dissipate. The pulse and pause duration (1-200ms and 4-5s, respectively), the pulse current density (40-120Acm$^{-2}$) or pulse potential (0.4-0.8V), the ion concentration (see above), the electrolyte convection (mechanical stirrer) and the temperature (room temperature) is set in order to deposit both, bismuth and telluride rich thermolegs [4] (Fig. 3). With shorter pulses, the material becomes denser, but at the same time the material's internal stress increases. Hence, cracks appear and the stoichiometric control cannot be guaranteed over the whole deposition area. Due to the high aspect ratio of the thermopile holes, ion diffusion is dominant over convection, for which reason the pause duration has to be rather long. The deposition of Bi-rich $Bi_2Te_3$ is performed in electrolyte baths with $Bi_2O_3$ concentrations of 40-60mmol. The Te-rich $Bi_2Te_3$ is deposited using $Bi_2O_3$ concentrations of 20-40mmol.

### III.    MEASUREMENT SETUP

The manufactured µTEGs are tested and characterized with the same measurement setup as used for the Cu/Ni generators [1]. The generators are clamped between a Peltier cooling element mounted on a water cooling system and three thermoresistors integrated in a metal bar. The temperature difference is measured and controlled by two thermoresistive sensors close to the generator contact area. The obtained potential difference and the internal electrical resistance of the µTEGs were measured in a four probe configuration. The matched load case maximum output power was subsequently calculated. Due to the limited thickness of the device and the non perfect thermal contact to heat source and sink, the temperature difference across the generator surfaces $\Delta T_G$ is significantly lower than the temperature difference measured between the sensors $\Delta T_{meas}$. The fraction of $\Delta T_{meas}$ that is lost in the interface is directly depending on the thermal interface resistance $K$, the TC length and the ratio of insulating to thermoactive area, $A_V$. The specific electrical contact resistance $\rho_C$ directly influences the power output by influencing the internal resistance $R_i$ of the generator.

### IV.    ANALYTICAL MODEL

The model based analysis [1] of the performance of the µTEGs is shown in Fig. 4. The characteristics are calculated for a $\Delta T_{meas}$ of 40K and the average $K$=3.9KW$^{-1}$ for Ni/Cu and $Bi_2Te_3$ generators for comparison. The optimal TC length for $Bi_2Te_3$ generators is around 100 - 300µm. This fits well to the capabilities of microsystems technology, applying SU-8 as photosensitive resist and flexible substrate material.

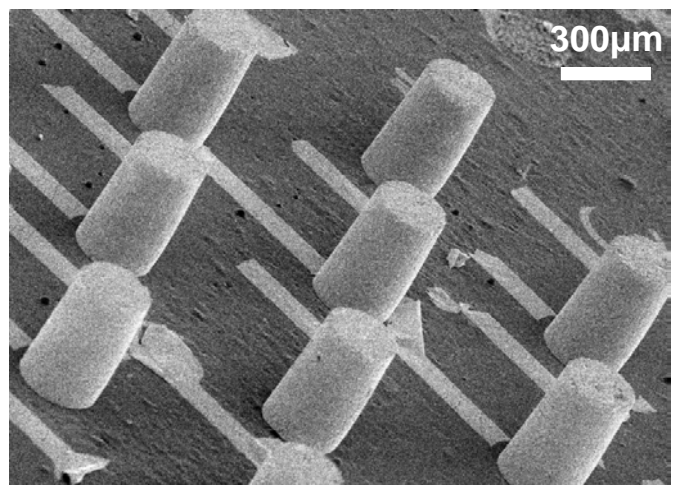

**Fig 3** Electrochemically deposited $Bi_2Te_3$ thermolegs with high aspect ratios.

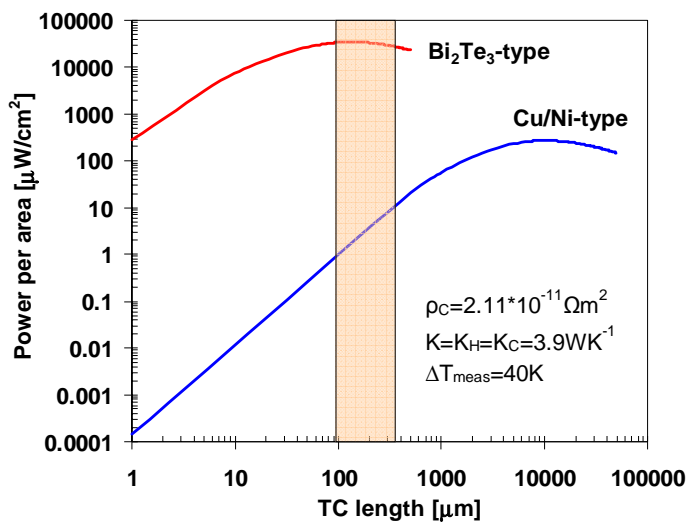

**Fig 4** Comparison of the performance of Cu/Ni and $Bi_2Te_3$ based µTEGs with double logarithmic axis.

## V. Measurement Results and Discussion

### A. μTEG Chip Processing

The first set of $Bi_{2+x}Te_{3-x}$ μTEGs yielded 71.6±1.0μWcm$^{-2}$ for a $\Delta T_{meas}$ of 40.3±0.1K (see Fig. 6). After an annealing step for 18 hours at 200°C, the performance increased by a factor of 3.9 and the generator delivered 278.5±5.4μWcm$^{-2}$ for a similar $\Delta T_{meas}$ of 40.0±0.3K. In comparison with the Cu/Ni generator, the $Bi_{2+x}Te_{3-x}$ generator is over 60 times better for the same operating conditions. The maximum output power measured so far was 344.1μWcm$^{-2}$ for $\Delta T_{meas}$=44.4K and results in an efficiency factor ZT of 0.18μWcm$^{-2}$K$^{-2}$.

### B. Electrochemical Deposition of $Bi_{2+x}Te_{3-x}$

As depicted in Fig. 4 and 5, thermolegs with constant stoichiometry and a thickness of over 300μm are deposited at a rate of 20μm/h. All the stoichiometric analysis have been made by Energy Dispersive X-ray Detection (EDX) and the measured stoichiometric ratios covering all the fabricated generators did range from 0.8 (Bi rich) to 2.1 (Te rich). Subsequent qualitative Seebeck measurements have reconfirmed that both, Bi-rich and Te-rich thermo legs with positive and negative Seebeck coefficients are achieved.

## VI. Conclusion

The fabrication of vertical, flexible $Bi_{2+x}Te_{3-x}$ μTEGs could be shown, by using a novel, fast and low cost ECD approach with deposition rates exceeding 20μm/h. This novel ECD method further more allows full control of the stoichiometry over the whole deposit. The deposition of n- and p-rich $Bi_{2+x}Te_{3-x}$ with a thickness of over 300μm and an aspect ratio of at least 1.5 could be proven. According to our analytical model, the stated process enables μTEG fabrication in the ideal thermoleg thickness regime. The μTEG performance could further be improved by a factor of 3.9, if thermal annealing was performed for 18h at 200°C.

In addition to annealing, we see several routes to optimization: adjust the stoichiometry of Te-rich and Bi-rich thermolegs for higher Seebeck coefficients by optimizing the process parameters during electroplating, reduce grain boundaries by adjusting the deposition speed, and reduce the internal resistance by improving contact and interconnect resistances. Electroplating allows increasing the thermo legs' length towards 200 - 300μm, which is the simulated optimum for the given material combination (see Fig. 3). Electroplating in general offers several advantages above sputtering of $Bi_{2+x}Te_{3-x}$ as thermoelectric material, which has yielded the current state-of-the-art μTEGs with an efficiency factor of 2.4μWcm$^{-2}$K$^{-2}$ [5]. These advantages are: high deposition speed and thicker layers, low material and energy consumption and low costs on large substrate sizes.

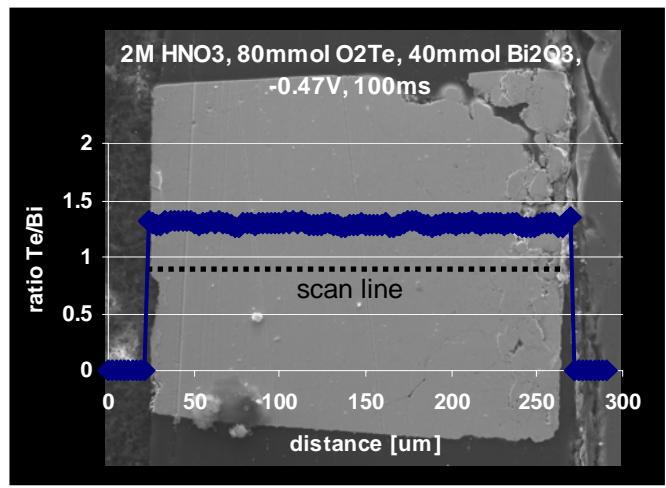

**Fig 5** SEM image with overlaid EDX line scan of an electrochemically deposited $Bi_{2+x}Te_{3-x}$ thermoleg measured in its growth direction.

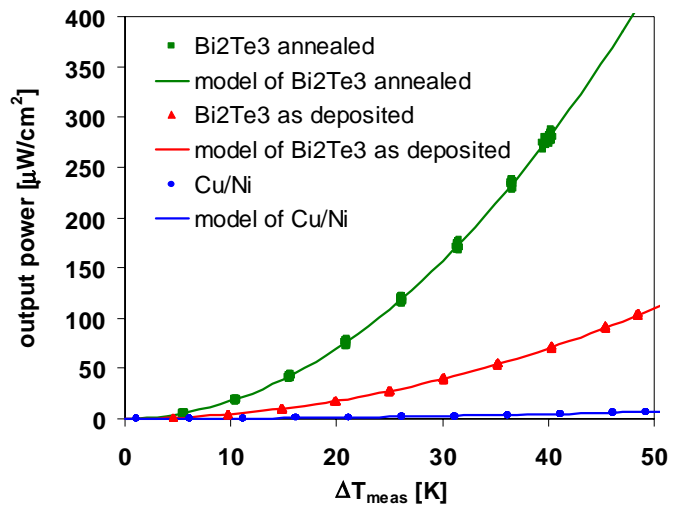

**Fig 6** Comparison of the measured matched output power for Cu/Ni, $Bi_2Te_3$ and annealed $Bi_2Te_3$ micro thermoelectric generators in comparison to the model based calculations.

### References

[1]  W. Glatz, S. Muntwyler, C. Hierold, „Optimization and Fabrication of Thick Flexible Polymer Based Micro Thermoelectric Generator", *Sensors and Actuators A-Physical, Vol. 132, pp. 337-345, 2006*

[2]  H. Takahashi, M. Kojima, S. Sato, N. Ohnisi, A. Nishiwaki, K. Wakita, T. Miyuki, S. Ikeda, and Y. Muramatsu, „Electric and thermoelectric properties of electrodeposited bismuth telluride (Bi2Te3) films", *Journal of Applied Physics, Vol. 96, pp. 5582-5587, 2004*

[3]  M. Martin-Gonzalez, A. Prieto, R. Gronsky, T. Sands, and A. Stacy, „Insights into the Electrodeposition of Bi2Te3", *Journal of the Electrochemical Society, Vol. 149 (11), pp. C546-C554, 2002*

[4]  W. Glatz, L. Durrer, E. Schwyter, C. Hierold, „Novel mixed method for the electrochemical deposition of thick layers of $Bi_{2+x}Te_{3-x}$ with controlled stoichiometry", *submitted to Chemistry of Materials, ACS Publications, November 2007*

[5]  H. Bottner, J. Nurnus, A. Gavrikov, G. Kuhner, M. Jagle, C. Kunzel, D. Eberhard, G. Plescher, A. Schubert, K.H. Schlereth, „New thermoelectric components using microsystem technologies", *J. Microelectromech. Syst., Vol. 13, pp. 414-420, 2004*